\documentclass{moriond}

\def\be{\begin{equation}}
\def\ee{\end{equation}}
\def\bea{\begin{eqnarray}}
\def\eea{\end{eqnarray}}

\usepackage{subcaption}
\usepackage{wrapfig}
\usepackage{hyperref}
\usepackage{graphicx}
\usepackage{xspace}

\newcommand{\lemaitre}{{\sc Lema\^{\i}tre}\xspace}
\newcommand{\calspec}{{\sc Calspec}\xspace}
\newcommand{\gaia}{{\sc Gaia}\xspace}

\begin{document}
\vspace*{4cm}
\title{Calibrating the cosmic distance ladder for independent dark energy constraints}

\author{ Enya Van den Abeele, on behalf of the LEMAITRE collaboration }

\address{CNRS IJCLab, 91405, Orsay, France}

\maketitle

\abstracts{
Recent cosmological results presented by the DESI collaboration show a clear deviation from the $\Lambda$CDM model when considering a time-dependent dark energy equation of state. This deviation is strongly driven by the SNe Ia surveys, which all share the same low redshift sample. The ZTF survey provides the current larger SNe Ia sample at low redshift, which will significantly increase the constraining power. In this context, the \lemaitre collaboration aims to provide new constraints on dark energy parameters using three independent SNe Ia datasets: ZTF, SNLS and HSC. This effort requires robust photometric calibration, with a particular focus on filter transmission measurement and star catalog calibration. The target for \lemaitre is to control the calibration systematics on distance moduli below 0.4\%.}

\section{Introduction}
The DESI collaboration latest results on cosmological parameters combining BAO, CMB and SNe Ia constrain show a tension between 2.8$\sigma$ and 4.2$\sigma$ with $\Lambda$CDM \cite{desi2025}. These results are strongly driven by the SNe Ia datasets, which share the same sample at low redshift, causing a $\sim$4\% deviation between the distance modulus and the DESI best fit residuals at redshifts between $z$=0.01 and $z$=0.1 \cite{regnault2026}. The advent of the Zwicky Transient Facility (ZTF) and its statistical power (3628 light curves published in the DR2 \cite{rigault24}) will allow to anchor the Hubble diagram at low redshift. The ZTF data motivated a new project called \lemaitre \cite{regnault2026}, proposing a new analysis of dark energy parameters by combining the ZTF dataset with nearly 200 SNe Ia from the Superova Legacy Survey (SNLS) and 300 from Hyper Suprime Cam (HSC). This analysis requires a redundant and robust calibration chain: as the SNe Ia spectra are resulting from ground based optical telescopes observations, they are extended on several filters passbands, and shift accross those passbands with the redshift, highlighting the need for precise filter transmission measurement. In a second step, those SNe Ia fluxes are compared with primary standard fluxes measured outside of the atmosphere (\calspec \cite{calspec}), and also by SNLS and ZTF. As those reference stars catalog are the link between the observed SNe Ia lightcurves and cosmology, SNLS and ZTF field stars can be photometrically calibrated with respect to those primary standard fluxes. Based on the DESI best fitting model, the goal of the \lemaitre calibration chain is to control systematic uncertainties on distances to less than 0.4 \%, which requires the knowledge of the filter transmission to be less than 0.2 nm and the calibration of star catalogs to within 0.1\%. We present in section 2 the measurement of the instrument passbands. We then sketch in section 3 the calibration of the photometric catalogs established independently by SNLS and ZTF. Lastly we discuss succinctly the consistency between the two surveys.

\section{Instrument model}\label{subsec:filters}


\subsection{ZTF}
\begin{figure}[h]
    \centering
    \begin{subfigure}[b]{0.49\textwidth}
        \centering
        \includegraphics[width=\textwidth]{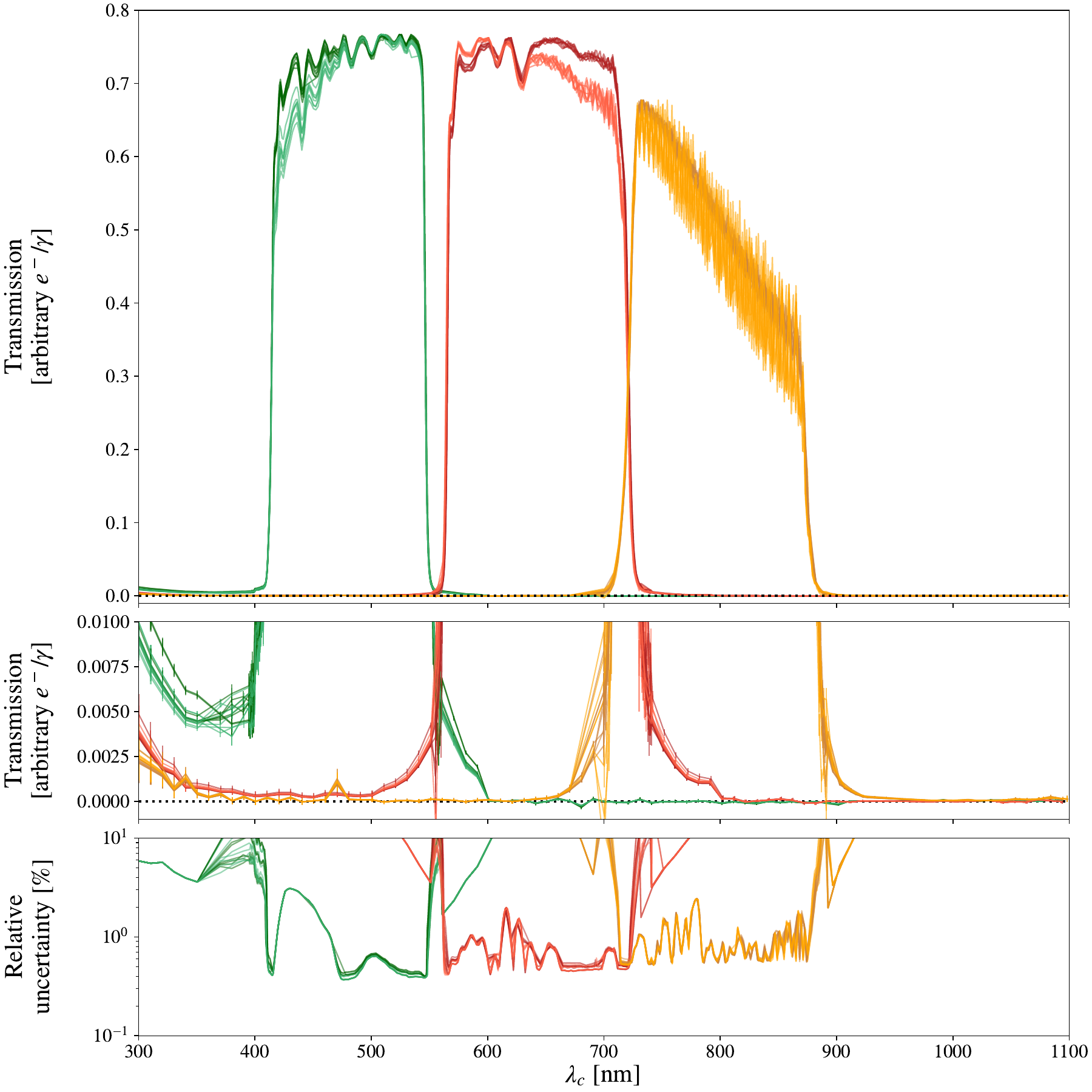}
        \caption{\fontsize{9}{12}\selectfont Top: ZTF filter transmission curves for each CCD as measured by the TCBP. Middle: Zoom illustrating out of band rejection and leakes. Bottom: Relative statistical uncertainty in the TCBP measurement.}
        \label{ztf_bandpasses}
    \end{subfigure}
    \hfill
    \begin{subfigure}[b]{0.40\textwidth}
        \centering
        \includegraphics[width=\textwidth]{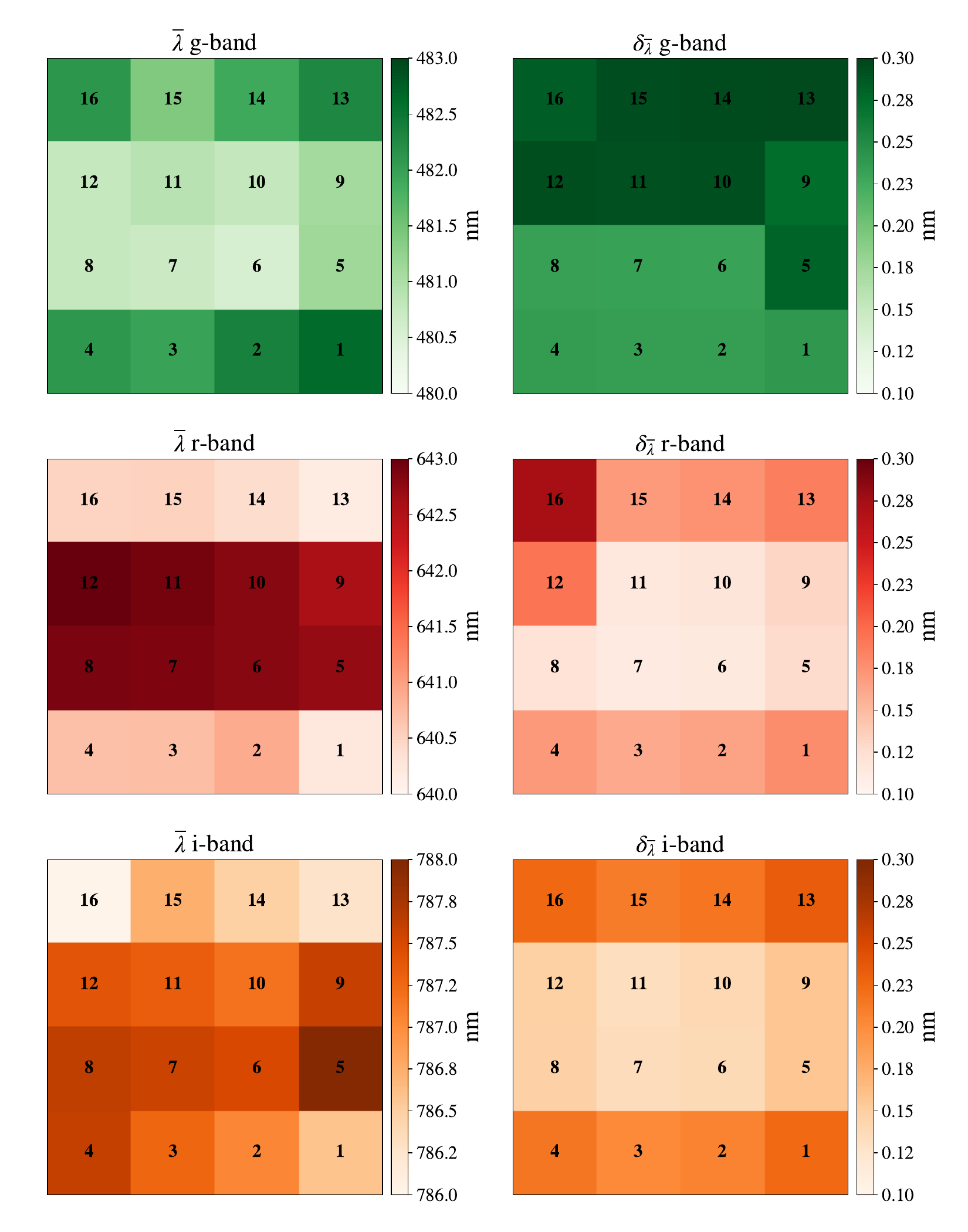}
        \caption{\fontsize{9}{12}\selectfont Mean lambda per CCD for each ZTF band $g$, $r$, $i$ (left column) and their associated uncertainties (right column).}
        \label{ztf_mean_lambdas}
    \end{subfigure}
\end{figure}

To determine the instrument model, it is necessary to accurately measure the transmission of the telescope and its filters. The ZTF telescope filters were calibrated using a Traveling Collimated Beam Projector (TCBP) \cite{kelian}, a device assembled with a white lamp and a telescope which projects a monochromatic parallel light beam. The flux of the light emitted by the TCBP  is monitored with a photodiode at the 0.01\% level, and the wavelength by a spectrograph at the 0.1 nm level. This instrument is the portable legacy of several predecessors, all sharing a similar operating principle \cite{coughlin} \cite{ingraham} \cite{souverin}.\\
A three-week campaign at the Palomar Observatory using the TCBP enabled several scans of the $g$, $r$, and $i$ bands with various samplings and different incident angles on the filters. The scanning strategy allowed to produce 16 transmission curves per filter, corresponding to one for each CCD, thereby accounting for variations in coating properties and quantum efficiency across the focal plane. The transmission curves are delivered with an uncertainty below 1\% for a resolution of 1 nm in the core of the passband (see Figure \ref{ztf_bandpasses}).\\
As a first result, the transmissions measured with the TCBP are showing strong differences with the public model, especially considering the structures within the core of the passbands and the position of the edges. Out-of-band leaks were also detected, especially one of 0.5\% below 400 nm in the $g$ band with a statistical significance above 20$\sigma$. \\
The computation of the mean wavelength per CCD and its uncertainty is taken as a metric to assess the precision of our filter measurement. Those mean wavelengths reveal variations up to 3 nm peak to peak across the focal plane in the three bands, which can already cause a color difference on the reference stars of about 10 milli-magnitudes if not taken into account. The mean wavelength uncertainties are below the nanometer level, up to 3 \r{A} in the $g$ band, and 2 \r{A} in $r$ and $i$ (see Figure \ref{ztf_mean_lambdas}). This result reaches the \lemaitre filter measurement requirement and makes ZTF the most accurately calibrated telescope in the Northern Hemisphere.

\subsection{SNLS}  
The MegaPrime instrument used by the SNLS survey also benefitted from in-situ measurement of its integrated measurement. The SnDICE calibration apparatus \cite{regnault2015} has a much lower wavelength resolution of 24 sampling points
covering the visible range, which is sufficient to probe the smooth transmission of the optics and detector quantum efficency, but insufficient to precisely determine the filter transmission. This in-situ measurement was complemented with a survey of filter transmissions conducted on the optical bench of the Laboratoire des Matériaux Avancés (LMA) \cite{Sassolas2018}. The combination result in a determination of the integrated passbands with a peak-to-valey uncertainty on central wavelength ranging between 0.4 and 0.6 nm depending on the band. The consistency of the model was tested using dense stellar fields observations to determine the color shift in the focal plane induced by variation in the transmission. The rms of the difference between \gaia colors from SNLS measurement and synthetic ones derived from the instrument model and main sequence star colors using the Pickles library spectra are dominated by the data photon noise. 

\section{Photometric calibration of the ZTF stars }

\begin{wrapfigure}{r}{8.cm}
\includegraphics[width=8.cm]{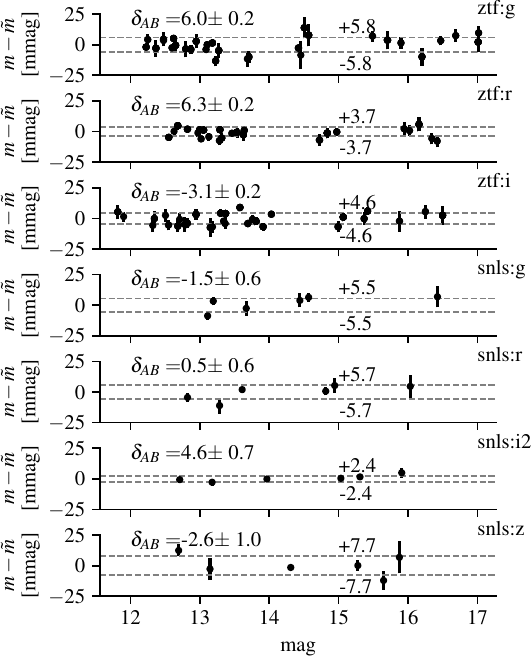}
\caption{\fontsize{9}{12}\selectfont Preliminary check of the internal consistency of ZTF and SNLS passbands. The $\delta_{AB}$ term represents the offset for \calspec magnitudes between synthetic ones derived from the instrument models and our calibrated aperture photometry catalogs. The rms for each band indicated in scattered lines are to be divided by the number of standard stars to get the final calibration uncertainty (Le Jeune et al., in prep.)}
\label{fig:deltaab}
\end{wrapfigure} 
The incident light from a star is altered in various ways before it reaches the telescope sensor: it goes through the atmosphere, telescope optics, detector and amplifiers. All of these factors affect the observed magnitude by an overall offset term known as the zero point (ZP), which evolves with time, airmass, or even the position on the focal plane. The \lemaitre calibration chain produces catalogs stars with a uniform ZP to serve as references for the photometry of supernovae. The absolute ZP is set by direct observations of \calspec primary standards. In ZTF, it is estimated using a photometric calibration method which takes advantage of numerous visits on the Northern Hemisphere sky: several exposures share the same stars observations, and this redundancy allows to fit the relative ZP. For the three ZTF bands, the light curves standard deviations show consistency across the entire footprint for more than one million stars, with an accuracy of less than 0.5\% for bright stars light curves. For fainter stars, the light curves residuals are dominated by Poisson noise. Then the relative ZP need to be anchored on an absolute flux scale, using the 30 \calspec stars present in the large ZTF footprint. For each band, the difference $\delta_{AB}$ between the aperture photometry catalog magnitudes and the synthetic ones determined using the ZTF instrument model is our link to cosmology. The final calibration uncertainty is below 0.1\% in the three filters, reaching the target of \lemaitre. \\
Regarding SNLS, to improve redundancy and provide calibration for the z band, dedicated direct observations of 6 \calspec stars were also performed with between 15 to 60 visits using the MegaPrime instrument. Despite the lower number of stars, the observations are sufficient to provide an independent anchor to the \calspec flux scale with a statistical uncertainty between 0.1 and 0.3 \%. Those results are shown in Figure \ref{fig:deltaab}. 


\section{ZTF-SNLS passband consistency}
Because the three \lemaitre surveys observe overlapping regions of the sky, their calibration chains include useful internal redundancies. For example, stars observed in common can be used to check the cross-consistency of the ZTF and SNLS passband models. This is done by looking at the \gaia stars shared in ZTF and SNLS fields, and use their spectra to determine synthetic magnitudes $\hat{m}^{ZTF}_b$ and $\hat{m}^{SNLS}_b$ considering the instruments models and calibrated catalogs. The difference between ZTF and SNLS measured magnitudes $m^{ZTF}_b$ and $m^{SNLS}_b$ in each band $b$ can be expressed as follows:
\begin{equation}
    m^{ZTF}_b - m^{SNLS}_b = \alpha_{meas}[m^g_{SNLS} - m^i_{SNLS}] + ZP
\end{equation}
where $\alpha$ is the color term and $m^g_{SNLS} - m^i_{SNLS}$ the $g-i$ color in SNLS, used as the variable. \\\\
After determining $\hat{m}^{ZTF}_b$ and $\hat{m}^{SNLS}_b$, the synthetic $\alpha_{syn}$ term is fitted using the \gaia stars spectra, and is then compared to the measured $\alpha_{meas}$ from the data. \\
The difference between $m^{ZTF}_b$ - $m^{SNLS}_b$ and $\hat{m}^{ZTF}_b$ - $\hat{m}^{SNLS}_b$ show a very good agreement with systematic error at the few per mil level for the three filter combinations $g_{ZTF} - g_{SNLS}$, $r_{ZTF} - r_{SNLS}$ and $i_{ZTF} - i_{SNLS}$.

\section*{Conclusion}

Within the \lemaitre pipeline, meticulous work has been done to measure the instruments transmission and to calibrate the photometric catalogs through a robust calibration chain. The filters mean wavelength uncertainties are below the nanometer level for the three surveys, and reach the 0.2 nm precision target for ZTF. Internal consistency and cross-consistency between ZTF and SNLS has shown very satisfying results with a final calibration uncertainty of 0.1\%, demonstrating the reliability and redundancy of this metrologic chain. \\
Future surveys like the Legacy Survey of Space and Time (LSST) will push the SNe Ia statistic even further with hundreds of thousands expected events, demanding to reduce the calibration uncertainties target below 0.1\%. The TCBP technique developed in the context of \lemaitre can be used to reach such a goal, as it demonstrated its efficiency on ZTF throughputs measurements.

\section*{Acknowledgement}
We are very grateful to the ZTF teams at Caltech and Palomar for their help and assistance during the TCBP campaign. We would also like to thank the Agence Nationale de la Recherche through the ANR-23-CE31-0007 project and DIM Origines (IDF-DIM-ORIGINES-2023-4-08) for their financial support.





\end{document}